\documentclass[a4paper,11pt]{article}
\pdfoutput=1 

\usepackage{jcappub} 

\usepackage[T1]{fontenc} 
\usepackage{natbib}
\usepackage{amsmath}

\title{\boldmath {\sc FAST-PT} : a novel algorithm to calculate convolution integrals in cosmological perturbation theory}

\author{Joseph E. McEwen,}
\author{Xiao Fang,}
\author{Christopher M. Hirata, and}
\author{Jonathan A. Blazek}

\affiliation{Center for Cosmology and AstroParticle Physics, Department of Physics, The Ohio State University, 191 W Woodruff Ave, Columbus OH 43210, USA}

\emailAdd{mcewen.24@osu.edu}

\abstract{We present a novel algorithm, {\sc FAST-PT}, for performing convolution or mode-coupling integrals that appear in nonlinear cosmological perturbation theory. The algorithm uses several properties of gravitational structure formation -- the locality of the dark matter equations and the scale invariance of the problem -- as well as Fast Fourier Transforms to describe the input power spectrum as a superposition of power laws. This yields extremely fast performance, enabling mode-coupling integral computations fast enough to embed in Monte Carlo Markov Chain parameter estimation. We describe the algorithm and demonstrate its application to calculating nonlinear corrections to the matter power spectrum, including one-loop standard perturbation theory and the renormalization group approach. We also describe our public code (in Python) to implement this algorithm. The code, along with a user manual and example implementations, is available at {\tt https://github.com/JoeMcEwen/FAST-PT}}

\newcommand{\dq}[1]{\frac{d^3 \mathbf{q}_{#1} }{(2 \pi)^3 } }
\newcommand{\Pl}{P_\text{lin}} 
\newcommand{\eqn}[1]{ \begin{equation} #1 \end{equation} } 
\newcommand{\Mpc}{\text{Mpc}}
\newcommand{\LegP}{{\cal P}}

\begin{document}
\maketitle
\flushbottom

\section{Introduction}
\label{sec:intro} 

The large-scale structure of the universe provides numerous probes of the underlying cosmological model, including the source of present-day accelerated expansion.
Current and upcoming surveys  \citep{Levi:2013gra,2013AJ....145...10D,2011arXiv1110.3193L,2016arXiv160100329D,2013arXiv1305.5422S} will provide impressive statistical power to test the $\Lambda$CDM (cosmological constant plus cold dark matter) paradigm as well as potential modifications (see Ref.~\cite{2013PhR...530...87W} for a review). Connecting the predictions of these models to observables from tracers of large-scale structure requires understanding the role of physics on a wide range of scales, including the growth of dark matter structure and the formation of galaxies and other luminous objects. On small scales, numerical simulations are required to solve for the full nonlinear growth (e.g.\ \cite{2006Natur.440.1137S}). Perturbative techniques provide an analytic approach to describe structure on mildly nonlinear scales and are particularly valuable in that they can be quickly calculated for different sets of cosmological parameters (without running a new simulation) and provide physical intuition into the relevant processes.

A generic feature of nonlinear perturbation theory is the coupling of modes at different scales through kernels that capture the physics of structure growth. As a result, these nonlinear corrections typically appear as convolutions over the power spectrum or related functions of the wavevector. In this paper, we primarily consider the most ubiquitous of these approaches, standard perturbation theory (SPT, e.g.\ \cite{2002PhR...367....1B}). However, integrals with a similar structure are found in other approaches as well, including Lagrangian perturbation theory (LPT, \cite{2014ApJ...788...63S}), renormalized perturbation theory (RPT, \cite{2006PhRvD..73f3519C}), renormalization group perturbation theory (RGPT, \cite{2007PhRvD..75d3514M,2014arXiv1403.7235M}, also considered in this work), the effective field theory (EFT, \cite{2012JCAP...07..051B, 2012JHEP...09..082C, 2013JCAP...08..037P, 2014PhRvD..89d3521H}) approach to structure formation, and time renormalization frameworks \cite{2011JCAP...10..037A}, which can include scale-dependent propagators for the fluctuation modes (e.g.\ arising from massive neutrinos). Therefore, it is of great utility that the cosmological community have access to efficient and accurate methods to compute these integrals.
 
The applicability of perturbative techniques is not limited to dark matter evolution.  A number of cosmological observables can be modeled in the weekly non-linear regime. These include the clustering of galaxies and other luminous tracers, as well as weak gravitational lensing and cross-correlations between these probes (e.g.\ ``galaxy-galaxy lensing.''). For instance, the relationship between dark matter and luminous tracers will generally include a nonlinear ``biasing'' relationship, resulting in correlations that are naturally described in a perturbative expansion (e.g.\ \cite{2006PhRvD..74j3512M, 2009JCAP...08..020M,2014PhRvD..90l3522S}). Many cosmological analysis limit their scope to the weakly non-linear regime, where the majority of the information is, and employ a bias expansion to constrain cosmological parameters \cite{2013MNRAS.432.1544M,2016arXiv160407871K} as well as, e.g., the total neutrino mass \cite{2011PhRvD..83d3529S,2013MNRAS.436.2038Z}. For instance, \S 2 of \cite{2016arXiv160407871K} demonstrates a recent application of nonlinear biasing. In the absence of a fast algorithm for performing the relevant convolutions, that work used emulation, calibrated with the results of a conventional method, to obtain the correct contributions at arbitrary cosmological parameters.

Perturbative techniques can predict the nonlinear shift and broadening of the baryon acoustic oscillation (BAO) feature \cite{2008PhRvD..77b3533C,2014JCAP...02..042S} -- a powerful ``standard ruler'' for studying the evolution of geometry in the universe -- including the potential impact of streaming velocities between baryons and dark matter in the early universe \citep{2011JCAP...07..018Y,2013PhRvD..88j3520Y,2015MNRAS.448....9S,2015arXiv151003554B}. The velocity field of dark matter and luminous tracers, which sources ``redshift-space distortions'' in clustering measurements can also be modeled analytically beyond linear theory (e.g.\ \cite{2004PhRvD..70h3007S,2012JCAP...11..009V}). Similarly, correlations of intrinsic galaxy shapes (known collectively as ``intrinsic alignments'') must be included in cosmic shear analyses and can be described perturbatively (e.g.\ \cite{2004PhRvD..70f3526H,2015JCAP...08..015B}). 
 
Although these examples indicate the broad applicability of perturbative techniques, some analyses will probe regimes where numerical simulations are required to reach the desired accuracy. Even in these cases, however, a fast perturbation theory code is still valuable, since interpolation (or emulation, \cite{2014ApJ...780..111H}) from grids of simulations can be used to compute the non-perturbative correction to an observable ${\cal O}$, rather than trying to interpolate the much larger ``raw'' value of ${\cal O}$.
 
In this paper we present {\sc FAST-PT}, a new algorithm and publicly available code to calculate mode coupling integrals that appear in perturbation theory. As a first example of our method we focus on 1-loop order perturbative descriptions of scalar quantities (e.g.\ density or velocity divergence). In particular, we present examples for 1-loop SPT, which can be trivially expanded to include nonlinear galaxy biasing, and renormalization group results. A generalization to arbitrary-spin quantities (e.g.\ intrinsic alignments, a spin-2 tensor field) and other directionally dependent power spectra (e.g.\ redshift-space distortions and secondary CMB anisotropies) will be presented in a follow-up paper \cite{Fang2016}.  

{\sc FAST-PT} can calculate the SPT power spectrum, to 1-loop order to the same level of accuracy as conventional methods, on a sub-second time scale. In the context of Monte Carlo Markov chain (MCMC) cosmological analyses, which may explore $>10^6$ points in parameter space, the extremely low recurring cost of our method is particularly relevant. The {\sc FAST-PT} recurring cost to calculate the 1-loop power spectrum at $N=3000$ $k$ values is $\sim 0.01$s. This speed is even more valuable for multi-probe cosmological analyses. For instance, a gravitational lensing plus galaxy clustering analysis may require the matter and galaxy power spectra in real and redshift space, nonlinear galaxy biasing contributions, and the intrinsic alignment power spectra, at each point in cosmological parameter space. {\sc FAST-PT} provides a means to obtain these quantities in a time that is likely trivial compared to other necessary calculations at each step in the chain (e.g.\ obtaining the linear power spectrum from a Boltzmann code).

{\sc FAST-PT} takes a power spectrum, sampled logarithmically, as an input. Special function identities are then used to rewrite the angular dependence of the mode-coupling kernels in terms of a summation of Legendre polynomials. The angular integration for each of these components can be performed analytically, reducing the numerical evaluation to one-dimension. Because of the uniform (logarithmic) sampling we are able to utilize Fast Fourier Transform (FFT) methods, thus enabling computation of the mode-coupling integrals in ${\cal O}(N\log N)$ operations, where $N$ is the number of samples in the power spectrum. Our approach is similar in structure to the evaluation of logarithmically sampled Hankel transforms \cite{talman1978numerical,2000MNRAS.312..257H}, which have been used to transform power spectrum into correlation functions (and vice versa). It also draws on the realization that convolution integrals in spherical symmetry -- even convolutions of integrands with spin -- can be expressed using Hankel transforms with the angular integrals performed analytically (e.g.\ \cite{2012PhRvD..85d3523F,2015MNRAS.448....9S}). We implement the {\sc FAST-PT} algorithm in a publicly-available package. The code is written in Python, making use of {\sc numpy} and {\sc scipy} libraries, and has a self-contained structure that can be easily integrated into larger packages. We provide a public version of the code along with a user manual and example implementations at {\tt https://github.com/JoeMcEwen/FAST-PT}. 

Recently, Schmittfull {\slshape et al.}~\cite{arXiv:1603.04405} have presented a related method for fast perturbation theory integrals, based on the same mathematical principles. Our Eq.~(\ref{J_k_2}) encapsulates the same approach as their Eq. (31), combined with the logarithmically sampled Hankel transform. However, the numerical approach is different: the decomposition of an arbitrary power spectrum $P(k)$ into power laws of complex exponent is treated as fundamental (and is kept explicitly in the code); the near-cancellation of $P_{22}+P_{13}$ is handled by explicit regularization; and the $P_{13}$ integral is solved using a different method (based only on scale invariance). Finally, we present a fast implementation of RGPT.

This paper is organized as follows: in $\S 2$, we provide the theory for our method, motivating the approach by considering the 1-loop SPT power spectrum. In $\S 3$, we provide results for 1-loop corrections to the power spectrum and demonstrate an implementation of the renormalization group approach of \cite{2007PhRvD..75d3514M,2014arXiv1403.7235M}. In $\S 4$, we summarize our results, including a discussion of other potential applications of {\sc FAST-PT}, and provide a brief description of the publicly-available code. The appendices provide additional details of our numerical calculations and the mathematical structure of the terms under consideration.

\section{Method}
\label{method_intro} 

This work presents an algorithm to efficiently calculate mode-coupling integrals of the form
\begin{align}
\label{eq:conv}
 \int \dq{} K(\mathbf{q}, \mathbf{k}- \mathbf{q}) P(q) P(|\mathbf{k}- \mathbf{q}|) ~, 
\end{align} 
where $K(\mathbf{q}_1, \mathbf{q}_2)$ is a mode-coupling kernel that can be expanded in Legendre polynomials and $P(q)$ is an input signal logarithmically sampled in $q$. The motivation for this method is mildly-nonlinear structure formation in the universe, although it can be more generally considered as a technique to evaluate a range of expressions in the form of Eq.~(\ref{eq:conv}). 

For clarity we list our conventions and notations:
\begin{itemize}
\item{fast Fourier transform and inverse fast Fourier transform are denoted as FFT and IFFT;}
\item{Fourier transform pairs have the $2\pi$ placed in the denominator of the wavenumber integral, as is standard in cosmology:
\begin{align}
\Phi(\mathbf k) = \int d^3\mathbf r\, \Phi(\mathbf x)\,e^{-i\mathbf k\cdot\mathbf r}
~~~\leftrightarrow~~~
\Phi(\mathbf r) = \int \frac{d^3\mathbf k}{(2\pi)^3} \, \Phi(\mathbf k)\,e^{i\mathbf k\cdot\mathbf r};
\end{align}
}
\item{ ``$\log$'' always refers to natural log and we will use $\log_{10}$ explicitly when we are referring to base 10;}
\item{ $\otimes$ represents a convolution (discrete or continous);}
\item{the Legendre polynomials will be denoted $\LegP_l$ (to avoid confusion with power spectra $P$), normal Bessel functions of the first kind are denoted $J_\mu(t)$, and spherical Bessel functions of the first kind are denoted $j_l(t)$, all with standard normalization conventions \cite{abramowitz1964handbook};}
\item{ $i = \sqrt{-1}$ (never used as an index);}
\item{``log sampling'' means that the argument of the input signal is $q_n=q_0 \exp(n \Delta)$, where $n=0,1,2, ...$ and $\Delta$ is the linear spacing between grid points;}
\item{we use the convention that when calculations require discrete evaluations, for example as in the case of discrete Fourier transforms, we index our vectors, while when evaluations are performed analytically we omit the index.}
\end{itemize}

In this section we begin by reviewing SPT (\S\ref{pt_review}); the reader who is already experienced with SPT may skip directly to $\S$\ref{method}. \S\ref{method} describes our main result: a rearrangement of the mode-coupling integral that allows $P_{22}$ and related integrals to be computed in order $N\log N$ operations. The $P_{13}$ integral is simpler than $P_{22}$, but brute-force computation of $P_{13}$ is in fact slower than the {\sc FAST-PT} method for $P_{22}$, so we describe our fast approach to $P_{13}$ in \S\ref{p13_sec}. Finally, in \S\ref{regularization} we describe our numerical treatment of the cancellation of infrared divergences in $P_{22}$ and $P_{13}$.

\subsection{1-loop Standard Perturbation Theory}
\label{pt_review}
When fluctuations in the density field are small, $\delta(k) \ll 1$, non-linear structure formation in the universe can be modeled by solving the cosmological fluid equations using perturbation theory (see Ref.~\cite{2002PhR...367....1B} for a comprehensive review of Eulerian perturbation theory). For this paper we only sketch out the most basic elements of perturbation theory, focusing on the integrals we evaluate. The matter field written as a perturbative expansion in Fourier space is
\begin{align} 
\delta(\mathbf{k})= \delta^{(1)}(\mathbf{k}) + \delta^{(2)}(\mathbf{k}) + \delta^{(3)}(\mathbf{k}) + ... ~, 
\end{align} 
where the first order contribution $\delta^{(1)}(\mathbf{k})$ is the linear matter field and each higher-order term represent non-linear contributions. Non-linear effects manifest themselves as mode-couplings in Fourier space, consequently each $\delta^{(n)}(\mathbf{k})$ is a convolution integral over $n$ copies of the linear field $\delta^{(1)}(\mathbf{q})$ with a kernel $F_n( \mathbf{q}_1, ..., \mathbf{q}_n)$: 
\begin{align} 
\delta^{(n)}(\mathbf{k}) = \int  \dq{1}... \dq{n} \delta^3_\text{D}( \mathbf{k}- \displaystyle \sum_{j=1}^n  \mathbf{q}_j) F_n(\mathbf{q}_1,..., \mathbf{q}_n) \delta^{(1)}(\mathbf{q}_1), ..., \delta^{(1)}(\mathbf{q}_n)~,
\end{align} 
where $\delta^3_\text{D}( \mathbf{k})$ is the three-dimensional Dirac delta function. The power spectrum $P(k)$ is defined as an ensemble average of the matter field $\delta(k)$, 
\begin{align}
\langle \delta(\mathbf{k})  \delta(\mathbf{k}') \rangle = (2 \pi)^3 \delta_\text{D}^3( \mathbf{k}+ \mathbf{k}') P(k) ~ .
\end{align} 
The first non-linear contribution to the power spectrum comes from ensemble averages taken up to $\mathcal{O}([\delta^{(1)}]^4)$:
\eqn{
\langle \delta(\mathbf{k}) \delta(\mathbf{k}') \rangle = \langle \delta^{(1)}(\mathbf{k}) \delta^{(1)}(\mathbf{k}') \rangle + \langle \delta^{(2)}(\mathbf{k}) \delta^{(2)}(\mathbf{k}') \rangle + 2\langle \delta^{(1)}(\mathbf{k}) \delta^{(3)}(\mathbf{k}') \rangle + ... ~,
}
which defines the one-loop power spectrum
\begin{align} 
\label{one_loop}
P_\text{1-loop}(k) = \Pl(k) + P_{22}(k) + P_{13}(k) ~,
\end{align} 
where $ \langle \delta^{(2)}(\mathbf{k}) \delta^{(2)}(\mathbf{k}') \rangle =(2 \pi)^3 \delta_\text{D}^3( \mathbf{k}+ \mathbf{k}') P_{22}(k)$ and $ 2\langle \delta^{(1)}(\mathbf{k}) \delta^{(3)}(\mathbf{k}') \rangle =(2 \pi)^3 \delta_\text{D}^3( \mathbf{k}+ \mathbf{k}') P_{13}(k)$.

\subsection{$P_{22}(k)$ type Convolution Integrals}
\label{method}

We first focus on $P_{22}(k)$, leaving the evaluation of $P_{13}(k)$ to a later subsection. $P_{22}(k)$ is a convolution integral that takes two copies of the linear power spectrum $\Pl(k)$ as inputs:
\begin{align}
\label{P_22} P_{22}(k) = 2 \int \dq{} \Pl(q) \Pl( | \mathbf{k} - \mathbf{q} | )  | F_2( \mathbf{q}, \mathbf{k}-\mathbf{q} ) |^2 ~.
\end{align} 
The $F_2$ kernel is 
\begin{align}
\begin{split}  
F_2( \mathbf{q}_1, \mathbf{q}_2 ) 
& =\frac{5}{7} + \frac{1}{2}  \mu_{12} \left( \frac{ q_1}{q_2} + \frac{q_2}{q_1} \right) + \frac{2}{7}\mu_{12}^2 
\\ &=
\frac{17}{21} \LegP_0( \mu_{12} ) + \frac{1}{2} \left( \frac{q_1}{q_2} + \frac{q_2}{q_2} \right)\LegP_1(\mu_{12}) + \frac{4}{21} \LegP_2( \mu_{12} )~,
\label{F_2}
\end{split} 
\end{align} 
where we have defined $\mu_{12}=\mathbf{q}_1\cdot \mathbf{q}_2/(q_1 q_2) = \hat{\mathbf q}_1\cdot\hat{\mathbf q}_2$, which is the cosine of the angle between $\mathbf q_1$ and $\mathbf q_2$.
Squaring this and substituting into Eq.~(\ref{P_22}), we find that the $P_{22}(k)$ power spectrum expanded in Legendre polynomials is 
\begin{align} 
\begin{split} 
\label{P_22_leg}
P_{22}(k) & = 2  \int \dq{1} \Big[  \frac{ 1219}{1470}   \LegP_0(\mu_{12}) + \frac{671}{1029}   \LegP_2(\mu_{12}) 
 +  \frac{32}{1715}  \LegP_4(\mu_{12}) + \frac{1}{3}  q_1^2 q_2^{-2} \LegP_2(\mu_{12})  \\
& \;\;\;  + \frac{62}{35}  q_1 q_2^{-1} \LegP_1(\mu_{12}) +  \frac{8}{35} q_1 q_2^{-1}\LegP_3(\mu_{12}) 
 +   \frac{1}{6} q_1^2 q_2^{-2} \LegP_0(\mu_{12})  \Big] \Pl(q_1) \Pl(q_2) ~,
\end{split} 
\end{align} 
where we have defined $\mathbf{q}_2 = \mathbf{k}- \mathbf{q}_1$ and used the $\mathbf q_1\leftrightarrow \mathbf q_2$ symmetry to combine terms. We note that the last Legendre component in Eq.~(\ref{P_22_leg}) will eventually lead to a formally divergent expression in the {\sc FAST-PT} framework. In \S\ref{regularization} we discuss this type of divergence (which can appear in other contexts) and explicitly show the cancellation.

Each Legendre component of Eq.~(\ref{P_22_leg}) is a specific case of the general integral
\begin{align}
 \label{J_k_1}
 J_{ \alpha \beta l }(k) = \int \dq{1} q_1^\alpha q_2^\beta \LegP_l(\mu_{12}) P(q_1) P(q_2) ~. 
 \end{align}
Note that we have now omitted the subscript ``lin'' on the power spectrum and carry on our calculations for a general input power spectrum. For SPT calculations the input power spectrum should be $\Pl(k)$, however there are cases when a general power spectrum input is required, such as renormalization group equations. Our method of evaluation draws on several key insights from the literature. The first is that the Legendre polynomial can be decomposed using the spherical harmonic addition theorem, and that in switching between real and Fourier space one may use the spherical expansion of a plane wave to achieve separation of variables; see the Appendix of Ref.~\cite{2015MNRAS.448....9S}. The second is the fast Hankel transform \cite{talman1978numerical,2000MNRAS.312..257H}. We also address a number of subtleties to make these ideas useful for the 1-loop SPT integrals.

\begin{figure}[tbp]
\label{fig:log_periodic}
\centering 
\includegraphics[width=.9\textwidth]{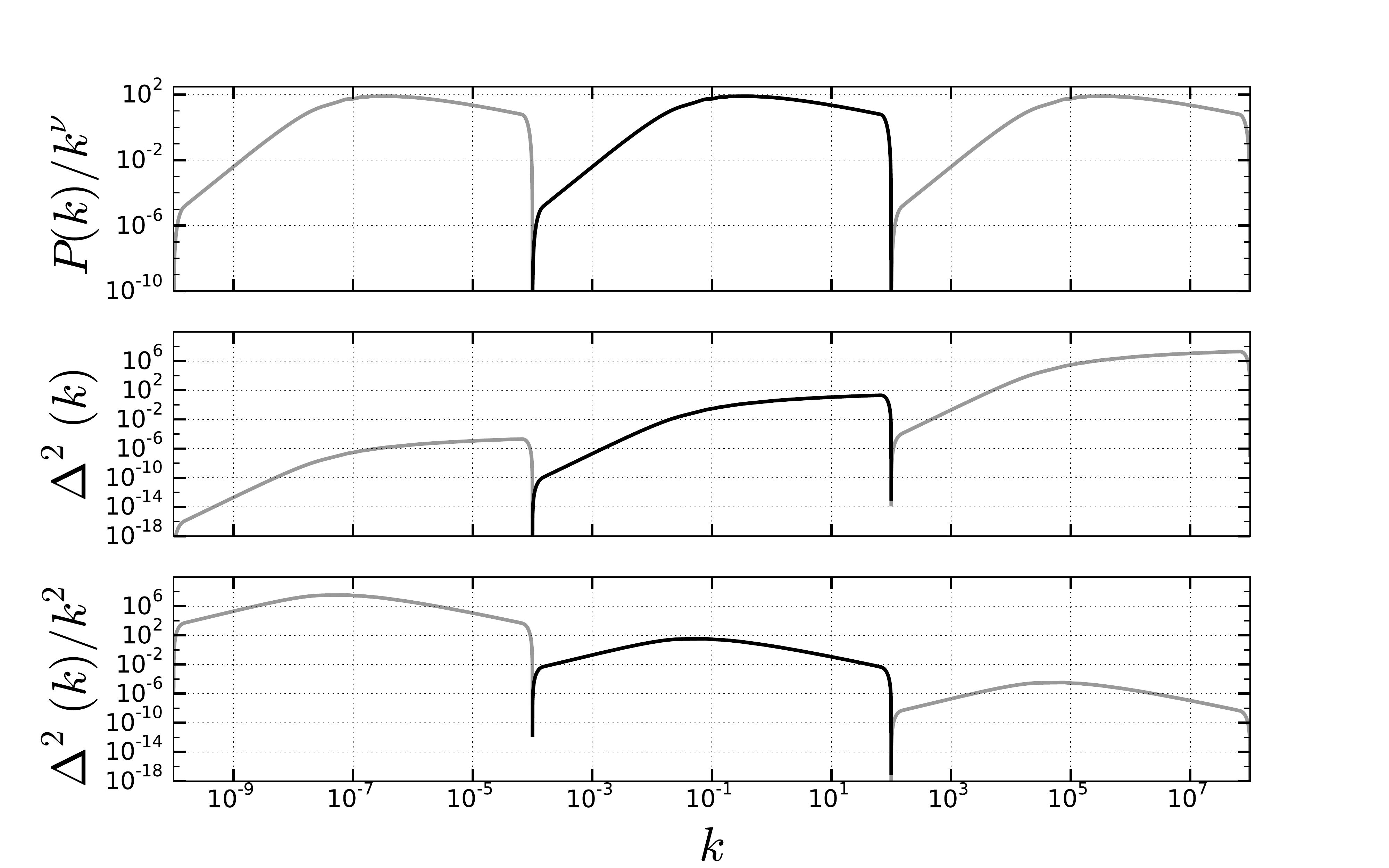}
\caption{\label{fig:sat} Power spectra in the log-periodic universe. Top panel shows the windowed linear power spectrum biased by $k^{-\nu}$ (we choose $\nu=-2$), with grey lines indicating the ``satellite" power spectra, i.e.\ the contribution to the total power spectrum that arises due to the periodic assumption in a Fourier transform. The middle panel plots $\Delta^2(k)=k^3P(k)/(2 \pi^2)$, within the periodic universe. This is the quantity that sources the density variance $\sigma^2= \int d \ln k \Delta^2(k)$. The bottom panel plots the contribution to the displacement variance $\sigma_\xi= \int d \ln k \Delta^2(k)/k^2$.}
\end{figure}

Our goal in this section is to develop an efficient numerical algorithm to evaluate integrals of the form Eq.~(\ref{J_k_1}). Combining the results for the relevant values of $(\alpha,\beta,l)$ will then allow us to construct $P_{22}(k)$ or other similar functions. For instance, in terms of these components, Eq.~(\ref{P_22_leg}) reads
\begin{align}
\begin{split} 
P_{22}(k)  & = 2 \Big[ \frac{1219}{1470}J_{0,0,0}(k) + \frac{671}{1029}J_{0,0,2}(k) + \frac{32}{1715}J_{0,0,4} (k) \\
& \;\;\; +\frac{1}{6}J_{2,-2,0}(k) + \frac{1}{3}J_{2,-2,2}(k) + \frac{62}{35}J_{1,-1,1}(k) + \frac{8}{35}J_{1,-1,3}(k) \Big].
\label{P22k-dec}
\end{split} 
\end{align}
To evaluate Eq.~(\ref{J_k_1}) we first Fourier transform to configuration space and then expand the Legendre polynomials in spherical harmonics, using  Eq.~(\ref{add_form}): 
\begin{align}
\begin{split} 
J_{\alpha \beta l}(r) &= \int \frac{d^3\mathbf{k}}{(2 \pi)^3} e^{i \mathbf{k} \cdot \mathbf{r}} J_{\alpha \beta l }(k) \\
& =  \int \dq{1} \dq{2}   e^{i (\mathbf{q}_2+ \mathbf{q}_1) \cdot \mathbf{r} } q_1^\alpha q_2^\beta P_l(\mu) P(q_1) P(q_2)\\
&=\frac{4 \pi}{2l + 1} \displaystyle \sum_{m=-l}^l   \int \dq{1} \dq{2}   e^{i \mathbf{q_1} \cdot \mathbf{r} }   e^{i \mathbf{q_2} \cdot \mathbf{r} } q_1^\alpha q_2^\beta 
Y_{lm}(\hat{\mathbf{q}}_1) Y^*_{lm}(\hat{\mathbf{q}}_2) P(q_1) P(q_2) ~.\end{split}
\end{align}
The $\mathbf q_1$ and $\mathbf q_2$ integrals can each be broken into a radial ($\int_0^\infty dq_1\,q_1^2$) and angular ($\int_{S^2} d^2\hat{\mathbf q}_1$) part; the angular parts do not depend on the power spectrum and can be evaluated analytically using Eq.~(\ref{angl_1}):
\begin{align}
\begin{split} 
J_{\alpha \beta l}(r)
&=\frac{4 \pi (4\pi\,i^l)^2}{(2\pi)^6(2l + 1)} \displaystyle \sum_{m=-l}^l  Y_{lm}(\hat{\mathbf{r}}) Y_{lm}^\ast(\hat{\mathbf{r}})
\int_0^\infty\! dq_1\,q_1^{2+\alpha} j_l(q_1 r) P(q_1)
\int_0^\infty\! dq_2\,q_2^{2+\beta} j_l(q_2 r) P(q_2).
\end{split}
\end{align}
Additionally we make use of the orthogonality relation, Eq.~(\ref{sph_ortho}), to eliminate the sum over $m$:
\begin{align}
\label{J_r_1} J_{\alpha \beta l}(r) = \frac{(-1)^l}{4 \pi^4} \left[ \int_0^\infty dq_1 q_1^{\alpha +2} j_l(q_1 r) P(q_1) \right]  \left[ \int_0^\infty dq_2 q_2^{\beta +2} j_l(q_2 r) P(q_2) \right]  ~.
 \end{align}
Equation~(\ref{J_r_1}) can be considered as one component of a correlation function. For instance, the correlation function $\xi_{22}(r)$ [the Fourier counterpart to $P_{22}(k)$] is built from Eq.~(\ref{J_r_1}) with the same $\alpha, \beta, l$ combinations and pre-factors as in Eq.~(\ref{P_22_leg}). Equation~(\ref{J_r_1}) is the product of two Hankel transforms (terms in brackets) with the relevant prefactor. We denote the bracketed terms in Eq.~(\ref{J_r_1}) as $I_{\alpha l}(r)$ and $I_{\beta l}(r)$.  To evaluate $I_{\alpha l}(r)$, we first take the discrete Fourier transformation of the power spectrum (biased by a power of $k$):
\begin{align}
c_m=\displaystyle W_m \sum_{n=0}^{N-1}  \frac{P(k_n)}{k_n^\nu} e^{- 2 \pi im n/N}
~~~\leftrightarrow~~~
P(k_n)=\displaystyle \sum_{m=-N/2}^{N/2} c_m k_n^{\nu + i \eta_m}~, 
\label{PknSum}
\end{align}
where  $N$ is the size of the input power spectrum, $\eta_m= m \times 2\pi/(N\Delta) $, $m=-N/2, -N/2 + 1,...,N/2-1, N/2$ and $\Delta$ is the linear spacing, i.e.\ $k_n=k_0\exp(n \Delta)$. For real power spectrum the Fourier coefficients obey $c_m^\ast=c_{-m}$. Here $W_m$ is a window function that can be used to smooth the power spectrum.\footnote{If no smoothing is desired, we would set $W_m=1$ for all $m$ except for $W_{\pm N/2}=\frac12$. The $\frac12$ ensures that the counting of {\em both} $m=\pm N/2$ in the second sum in Eq.~(\ref{PknSum}) is the correct inverse transform. However, in our numerical implementation we always include a window function that goes smoothly to zero to prevent ``ringing'' in the interpolated $P(k)$; see Appendix~\ref{sec:edge}.} Using discrete FFTs allows a significant reduction in computation time. However, these methods require that the function being transformed is (log-)periodic. In the case of {\sc FAST-PT}, this procedure is equivalent to performing calculations in a universe with a power spectrum, biased by a power-law in $k$, that is log-periodic. This universe has divergent power on large or small scales, depending on the choice of $\nu$. Figure~\ref{fig:log_periodic} shows the resulting power spectrum, with a window function applied at the periodic boundaries. In order for perturbation theory to make sense, the large-scale density variance (i.e.\ $\int_0^k k'{^2}P(k') dk'$) and the small-scale displacement variance (i.e.\ $\int_k^\infty P(k')\,dk'$) should both be finite (see Fig.~\ref{fig:log_periodic}). Since $P(k)/k^\nu$ is log-periodic, this means that {\sc FAST-PT} will require biasing with $-3<\nu<-1$ (this paper chooses $\nu=-2$).

In most cases, sufficiently far from the boundaries, the impact of the periodic nature of the $P(k)$ is negligible. However, while $P_{22}(k)$ and $P_{13}(k)$ are well-behaved in standard methods with CDM power spectra, they are both infinite in {\sc FAST-PT} where the satellite features at extremely large scales ($k\rightarrow 0$) produce infinite displacements. This is the same infinity found in power-law spectra and is of no physical concern: since displacement is not a physical observable, Galilean invariance guarantees that the divergent parts of $P_{22}(k)$ and $P_{13}(k)$ will cancel as long as the displacement gradient or strain is finite. In \S\ref{regularization} we will address the numerical aspects of this cancellation and show how to perform a well-behaved 1-loop SPT calculation in the {\sc FAST-PT} framework.

Continuing our evaluation:
\begin{align} 
\label{I_int}
\begin{split} 
I_{\alpha l }(r) &= \int_0^\infty dk\, k^{\alpha +2} j_l(k r) P(k) \\
& = \displaystyle \sum_{m=-N/2}^{N/2} c_m \int_0^\infty dk\, k^{\nu + 2 + \alpha + i \eta_m} j_l(kr) \\
& = \sqrt{\frac{\pi}{2}} \displaystyle \sum_{m=-N/2}^{N/2} c_m r^{-3 - \nu - \alpha - i \eta_m} \int_0^\infty dt  \; t^{3/2 + \nu  + \alpha + i \eta_m} J_{l + 1/2}(t)  \\
& =\sqrt{ \frac{ \pi }{2} }\displaystyle \sum_{m=-N/2}^{N/2} c_m g_{\alpha m} r^{-3-\nu-\alpha- i \eta_m}2^{Q_{\alpha m}} ~,
\end{split} 
\end{align}
where in the third equality we have exchanged the Bessel function of the first kind for a spherical Bessel function, $j_\nu(z) = \sqrt{\pi/(2z)}\,J_{\nu+1/2}(z)$ and performed the substitution $t = kr$. In the last equality we have evaluated the integral according to Eq.~(\ref{TSJ}) and defined $g_{\alpha m} \equiv g(l+\frac12,Q_{\alpha m})$ and $Q_{\alpha m}\equiv \frac32+\nu+\alpha+i\eta_m$.

%
%
%

Strictly speaking, the convergence criteria for Eq.~(\ref{I_int}) are $\alpha<-1-\nu$ and $\alpha+l >-3-\nu$. For $\nu=-2$ we thus require (i) $\alpha<1$ and (ii) $\alpha+l>-1$. All terms with $\alpha=2$ violate (i), while the $\alpha=2$, $l=0$ term also violates (ii). The violations of condition (i) can be cured if we apply an exponential cutoff in the power spectrum to force the integral to converge, i.e.\ in Eq.~(\ref{I_int}) we insert a factor of $e^{-\epsilon k}$ and take the limit as $\epsilon\rightarrow 0^+$; this yields the same result and is equivalent to smoothing out the ``wiggles'' in the Bessel functions at high $k$.\footnote{This can be proven by inserting a factor of $e^{-\epsilon t}$ in Eq.~(\ref{TSJ}) and taking the limit as $\epsilon\rightarrow 0^+$. Following Eq.~(6.621.1) of \cite{1994tisp.book.....G}, the integral can be expressed in terms of the hypergeometric function $_2F_1(\frac{\mu+\kappa+1}2,\frac{\mu+\kappa+2}2;\mu+1;-\epsilon^{-2})$. The transformation formula, Eq.~(9.132.2) of \cite{1994tisp.book.....G}, can then be used to express a hypergeometric function of large argument $-\epsilon^{-2}\rightarrow-\infty$ in terms of functions of argument approaching 0. Using $\lim_{z\rightarrow 0}\,_2F_1(\alpha,\beta;\gamma;z)=1$ and the $\Gamma$-function duplication formula suffices to prove this generalized version of Eq.~(\ref{TSJ}).} The violation of condition (ii) comes from the low $k$'s and is more problematic: the physical result for $I_{-2,0}(r)$ is divergent, and this will be treated in \S\ref{regularization}.
The final result for the $J_{\alpha \beta l}(r)$ correlation component is then
\begin{align} 
\label{J_r_2}
\begin{split} 
J_{\alpha \beta l}(r)
& = \frac{(-1)^l}{4 \pi^4} I_{\alpha l}(r)  I_{\beta l}(r)  \\
&=  \frac{(-1)^l}{8 \pi^3}  \displaystyle \sum_{m=-N/2}^{N/2} \displaystyle \sum_{n=-N/2}^{N/2} c_m c_n g_{\alpha m} g_{\beta n}\,2^{Q_{\alpha m}+Q_{\beta n}} r^{-6-2\nu-\alpha-\beta-i\eta_m-i\eta_n}~.  
\end{split}
\end{align} 
To obtain the power spectrum, we Fourier transform Eq.~(\ref{J_r_2}) back to $k$-space:
\begin{align}
\begin{split} 
J_{\alpha \beta l}(k_q) & = \int_0^\infty dr \,4 \pi r^2 j_0(k_qr) J_{\alpha \beta l}(r) \\
& = \frac{(-1)^l}{2 \pi^2} \displaystyle \sum_{m=-N/2}^{N/2} \displaystyle \sum_{n=-N/2}^{N/2} c_m g_{\alpha m} c_n g_{\beta n} 2^{Q_{\alpha m} +Q_{\beta n}}  \int_0^\infty dr\, r^{-5 -2 \nu - \alpha -\beta - i( \eta_m + \eta_n)} \frac{\sin(k_q r)}{k_q} ~,
\end{split}
\end{align}
where in the first equality homogeneity converts the 3-dimensional Fourier transform into a Bessel integral, and then we have used $j_0(z)= \sin(z)/z$.  The integral over $r$ can be evaluated using the $f$-function of Eq.~(\ref{gamma_sin_eqn}) via the substitution $t=k_qr$ and leads to
\begin{align}
\begin{split} 
\label{J_k_rk}
J_{\alpha \beta l}(k_q) = \frac{(-1)^l}{2 \pi^2} \displaystyle \sum_{m=-N/2}^{N/2} \displaystyle \sum_{n=-N/2}^{N/2} c_m g_{\alpha m} c_n g_{\beta n} 2^{Q_h} k_q^{-p-2+ i \tau_h} f_h~,
\end{split} 
\end{align}
where we have defined $f_h=f(p+1-i\tau_h)$, $p=-5-2\nu-\alpha-\beta$, $\tau_h= \eta_m + \eta_n$, and $Q_h = Q_{\alpha m}+Q_{\beta n}$. Note that $\tau_h$ (and hence $f_h$) and $Q_h$ depend only on the sum $m+n$.

In what follows, we will transform a double summation over $m$ and $n$ into a discrete convolution, indexed by $h$, such that $h=m + n\in\{-N,-N+1,...,N-1,N\}$. This leads to:
\begin{align}
\begin{split} 
\label{J_k_2}
J_{\alpha \beta l}(k_q) & = \frac{(-1)^l}{ 2\pi^2} 2^{3 +2\nu +\alpha + \beta} \displaystyle \sum_{m=-N/2}^{N/2} \displaystyle \sum_{n=-N/2}^{N/2} c_m g_{\alpha m} c_n g_{\beta n}\; f_h k_q^{-p-2 + i \tau_h} 2^{i \tau_h}  \\
& =  \frac{(-1)^l}{ \pi^2} 2^{2 +2\nu +\alpha + \beta}  \displaystyle \sum_h [c_m g_{\alpha m} \otimes c_n g_{\beta n}]_h f_h  2^{i \tau_h} k_q^{-p-2 + i \tau_h}  \\
&=  \frac{(-1)^l}{ \pi^2} 2^{2 +2\nu +\alpha + \beta} k_q^{-p-2} \displaystyle \sum_h C_h f_h  2^{i \tau_h} \exp(i \tau_h \log k_0 ) \exp( i \tau_h q\Delta)  \\
& =  \frac{(-1)^l}{ \pi^2} 2^{2 +2\nu +\alpha + \beta} k_q^{-p-2} \text{IFFT}[ C_h f_h  2^{i \tau_h} ]~,
\end{split}
\end{align} 
where in the second equality we have replaced $n=h-m$ and in the third and fourth equality the sum over $m$ is written as a discrete convolution $ \sum_m c_m g_{\alpha m} c_{h-m} g_{ \beta, h-m} = [c_m g_{\alpha m} \otimes c_n g_{\beta n}]_h= C_h $.  Also, due to the log sampling of $k_q$ the final sum over $h$ in Eq.~(\ref{J_k_2}) is actually an inverse discrete Fourier transform, i.e.\ $\sum_h A_h k_q^{i \tau_h}= \sum_h A_h\exp(i 2 \pi h q /[2N])$ \footnote{In the last two lines of Eq.~(\ref{J_k_2}) a shift, $\exp(i \tau_h \log k_0 )$, in the Fourier transform appears. In practice, our code does not compute this shift which also appears in the initial Fourier transform and thus cancels. Additionally, to conform to Python Fourier conventions we drop the positive end point in the final FFT.}, and can thus be evaluated quickly using an FFT. Equation~(\ref{J_k_2}) is the main analytical result of this work, it allows one to evaluate $P_{22}(k)$ type integrals quickly, scaling with $N\log N$. 

Since in {\sc FAST-PT} $P(k)/k^\nu$ is log-periodic, there are discontinuities in the power spectrum at $k_{\rm min}=k_0$ and $k_{\rm max} = k_0e^{N\Delta}$. This means that when Fourier-space methods are applied, the series of Eq.~(\ref{PknSum}) will exhibit ringing; the {\sc FAST-PT} user has several options for controlling this behavior. The power spectrum can be windowed in such a way that the edges of the array are smoothly tapered to zero (of course, this must be done outside the $k$-range that contributes significantly to the mode-coupling integrals). The location of the onset of the tapering is controlled by the user. The Fourier coefficients $c_m$ can also be filtered so that the highest frequencies are damped.
We use the same window function to filter the Fourier coefficients and smooth the edges of the power spectrum -- the functional form is presented in Appendix~\ref{sec:edge}. In practice, while we always apply a filter to the $c_m$ coefficients, we choose to directly window the power spectrum only within our renormalization group routine (see Appendix~\ref{rg_int}). We have also written the code in such a way that the user can easily implement their own window function. One can also ``zero pad'' the input power spectrum, adding zeros to both sides of the array. The contributions of the mode-coupling integrals from the large-scale satellite power spectrum ($k<k_{\rm min}$) heavily contaminate $P_{22}(k)$ at $k<2k_{\rm min}$ (range restricted by the triangle inequality). We thus recommend zero-padding by a factor $\ge 2$.

\subsection{$P_{13}(k)$ type Convolution Integrals}
\label{p13_sec}
The $P_{13}(k)$ integral does not share the same form as $P_{22}(k)$, since the wavenumber structure is different: it describes a correction to the propagator for Fourier mode $\mathbf k$ due to interaction with all other modes $\mathbf q$. The structure of $P_{13}(k)$ is thus $P(k)$ times an integral over the power in all other modes:
\begin{align}
\label{P13_Z}
P_{13}(k) = \frac{k^3}{252(2 \pi)^2} \Pl(k) \int_0^\infty dr\,  r^2 \Pl(kr) Z(r) ~, 
\end{align} 
where 
\begin{align} 
Z(r) = \frac{12}{r^4} - \frac{158}{r^2} + 100 - 42r^2 + \frac{3}{r^5}(7r^2 + 2)(r^2-1)^3 \log \frac{r+1}{| r-1|} ~,
\end{align}
and $r=q/k$. Upon making the substitution $r=e^{-s}$, Eq.~(\ref{P13_Z}) becomes 
\begin{align}
\begin{split} 
P_{13}(k) & = \frac{k^3}{252(2 \pi)^2 }\Pl(k) \int_0^\infty dr \; r^2 \Pl(kr) Z(r) \\
& = \frac{k^3}{252(2 \pi)^2 }\Pl(k) \int_{- \infty}^\infty  ds \; e^{-3s}  \Pl(e^{\log k - s})  Z(e^{-s}) \\
& = \frac{k^3}{252(2 \pi)^2 }\Pl(k)   \int_{- \infty}^{\infty} ds \; G(s) F(\log k - s)  ~,
\end{split} 
\end{align} 
where in the final line we reveal the integral as a continuous integral with the following definitions $G(s)\equiv e^{-3s}Z(e^{-s})$ and $F(s)\equiv \Pl(e^{s})$. In the discrete domain we have $ds \to \Delta$, $\log k_n=\log k_0+n\Delta$, and $s_m=\log k_0+m\Delta$, so that the discrete form is
\begin{align}
\begin{split}
\int_{- \infty}^{\infty} ds \; G(s) F(\log k - s)
& \rightarrow  \Delta \displaystyle \sum_{m=0}^{N-1}  G_D(m)  F_D(n-m) ~, 
\end{split}
\end{align}
where in the final line we define the discrete functions $G_D(m)\equiv G(s_m)$ and $F_D(m)\equiv F(m\Delta)$, so that we have
\begin{align}
\begin{split}
P_{13}(k_n) & = \frac{k_n^3}{252(2 \pi)^2 }\Pl(k_n) \Delta  [G_D \otimes F_D][n]~.
\end{split}
\end{align}
Thus $P_{13}(k)$, which at first appears to involve order $N^2$ steps (an integral over $N$ samples at each of $N$ output values $k_n$) can in fact be computed for all output $k_n$ in $N\log N$ steps.

\subsection{Regularization}
\label{regularization}

As mentioned above, we need to regularize the divergent portion in $P_{22}(k)$ with $P_{13}(k)$. In standard calculations in a $\Lambda$CDM universe, the suppression of power on large scales [$P(k)\propto k^n$, $n>-1$] controls this divergence, allowing the numerical evaluation of each term separately. The relevant cancellation will then occur upon addition of the terms, as long as sufficient numerical precision has been achieved. However, because the {\sc FAST-PT} method relies on FFTs, the ``true'' underlying power spectrum is log-periodic, leading to non-vanishing power on infinitely large (and small) scales. These divergences are thus numerically realized and must be analytically removed before evaluation. Physically the divergences are due to the artificial breaking of local Galilean invariance when the 1-loop SPT power is split into $P_{22}(k)$ and $P_{13}(k)$: a long-wavelength ($q\ll k$) velocity perturbation displaces small-scale structure without affecting its evolution, but since the perturbative expansion terms $\delta^{(n)}$ are defined with respect to a stationary background, each term in perturbation theory shows a divergence even when the physically relevant sum does not. This fact is well-known in the context of $P_{22}+P_{13}$ \cite{1983MNRAS.203..345V} and has been generalized to higher orders \cite{1996ApJ...456...43J,1996ApJS..105...37S}.

We construct our regularization scheme so that it preserves the 1-loop contribution to power spectrum, i.e. 
\begin{align}
 P_{22}(k) + P_{13}(k) =  P_{22,\text{reg}}(k) + P_{13,\text{reg}}(k) ~,
 \end{align}
where the subscript ``reg'' stands for regularization, by subtracting out the contribution to $P_{13}(k)$ from small $q=kr$ in Eq.~(\ref{P13_Z}), and adding it to the $J_{2,-2,0}(k)$ contribution in $P_{22}(k)$ to obtain a regularized $P_{22,\text{reg}}(k)$.  We first expand the kernel in Eq.~(\ref{P13_Z}) in a Laurent series around small $r$:
\begin{align}
\begin{split} 
r^2Z(r) = -168 + \frac{928}5r^2 - \frac{4512}{35}r^4 + \frac{416}{21}r^6 + \frac{2656}{1155}r^8 + ...
~.
\end{split} 
\end{align} 
If $P_{13}(k)$ were dominated by contributions from large-scale modes (i.e.\ $r\ll 1$), as occurs when there is an infrared divergence, then we could make the replacement $r^2Z(r)\rightarrow -168$ and find that $P_{13}(k)$ approaches
\begin{align}
\label{P_13_asym}
P_{13}(k) \rightarrow  - \frac{168 \; k^3}{252(2 \pi)^2 }\Pl(k) \int_0^\infty dr \Pl(kr)= - \frac{1}{3} k^2 \Pl(k) \int \dq{} \frac{\Pl(q)}{q^2} ~. 
\end{align} 
We then subtract this off from the kernel $Z(r)$ so that 
\begin{align} 
Z_\text{reg}(r) & = Z(r) + \frac{168}{r^2} =   \frac{12}{r^4}  + \frac{10}{r^2} + 100 - 42r^2 + \frac{3}{r^5}(7r^2 + 2)(r^2-1)^3 \log \frac{r+1}{| r-1|} ~.
\end{align}
The regularized version of $P_{13}(k)$ is
\begin{align} 
\begin{split} 
P_{13,\text{reg}}(k) & = \frac{k^3}{252(2 \pi)^2 }\Pl(k) \int_0^\infty dr \; r^2 \Pl(kr) Z_\text{reg}(r) \\
& = \frac{k^3}{252(2 \pi)^2 }\Pl(k) \int_{- \infty}^\infty  ds \; e^{3s}  \Pl(e^{\log k + s})  Z_\text{reg}(e^s) ~,
\end{split} 
\end{align} 
which can be evaluated numerically in the same manner that was presented in $\S$ \ref{p13_sec}. To regularize $J_{2,-2,0}(k)$ we take the power that we subtracted from $P_{13}(k)$ 
\begin{align}
\label{delta_pk}
\Delta P(k)= P_{13}(k) - P_{13,\text{reg}}(k) =  -\frac{k^2}{3} P(k) \int  \dq{} \frac{ \Pl(q)}{q^2}~, 
\end{align} 
and add it to $J_{2,-2,0}(k)$. To do this, we first take the Fourier transform of Eq.~(\ref{delta_pk}):
\begin{align}
\label{delta_xi}
\begin{split}
\Delta \xi(r)=
 \int \dq{1} e^{i \mathbf{q}_1 \cdot \mathbf{r}} \Delta P(q_1) & = -\frac{1}{3} \int \dq{1}  e^{i \mathbf{q}_1 \cdot \mathbf{r}} q_1^2 \Pl(q_1) \int \dq{2} \frac{\Pl(q_2)}{q_2^2} \\
& = -  \frac{1}{12 \pi^4}  \left[  \int_0^\infty dq_1\, q_1^4 \Pl(q_1) j_0(q_1 r) \right] \left[ \int_0^\infty dq_2\, \Pl(q_2) \right] ~.
\end{split} 
\end{align}
Since $J_{2,-2,0}(r)$ appears in $\xi_{22}(r)$ with a factor of $\frac13$ -- see Eq.~(\ref{P22k-dec}) -- it follows that $3\Delta\xi(r)$ should be added to $J_{2,-2,0}(r)$ if we want to preserve the sum $P_{22}(k)+P_{13}(k)$ in the regularization process. This leads to a regularized $J_{2,-2,0}(r)$:
\begin{align} 
\begin{split} 
\label{J_reg_r}
J_{[2,-2,0\; \text{reg}]}(r) &= J_{2,-2,0}(r) + 3\Delta \xi(r) \\
& =  \frac{1}{4 \pi^4} \left[ \int_0^\infty dq_1 \,q_1^4 \Pl(q_1) j_0(q_1 r) \right] \left\{ \int_0^\infty dq_2\,  \Pl(q_2) [ j_0(q_2 r)-1 ] \right\} ~. 
\end{split}
\end{align} 
The left bracket of Eq.~(\ref{J_reg_r}) proceeds in the same manner as presented in \S\ref{method}. The right bracket in Eq.~(\ref{J_reg_r}) requires some additional work:
\begin{align} 
\begin{split} 
I_{-2,0,\text{reg} } & = \int_0^\infty dq_2 \, \Pl(q_2) [ j_0(q_2 r)-1] \\
& = \displaystyle \sum_{n=-N/2}^{N/2}  c_n  \int_0^\infty dq_2 \, q_2^{\nu + i \eta_n}  \left[ \frac{\sin(q_2 r)}{q_2 r}-1\right] \\
& = \sum_{n=-N/2}^{N/2} c_n r^{-1-\nu-i \eta_n} g_n^\text{reg} ~, 
\end{split}
\end{align} 
where the integral may be evaluated by substituting $z=kr$ and finding:
\begin{align}
\begin{split} 
\label{g-reg-int}
g^\text{reg}_n(Q^\text{reg}_n) & =  \int_0^\infty dz\,  z^{\nu + i \eta_n}  \left( \frac{\sin z}{z}-1\right)
 =  \Gamma(Q^\text{reg}_n )\sin \frac{\pi Q^\text{reg}_n}{2}
 = f(Q^\text{reg}_n)~
\end{split}
\end{align}
and $Q^\text{reg}_n=  \nu +  i \eta_n$.\footnote{This integral is valid for its range of convergence, $-3<\nu<-1$. A straightforward way to prove this is to insert a factor of $e^{-\epsilon z}$, with $\epsilon$ small and positive, into the integrand; then expanding $\sin z=(e^{iz}+e^{-iz})/(2i)$ leads to a sum of three $\Gamma$-functions, two with $\Gamma(Q^\text{reg}_n)$ and one with $\Gamma(Q^\text{reg}_n+1)$. Taking the limit of $\epsilon\rightarrow 0^+$ causes the latter to drop out and the remaining two to give Eq.~(\ref{g-reg-int}).} The last equality uses Eq.~(\ref{gamma_sin_eqn}), and ensures that $g^\text{reg}_n(Q^\text{reg}_n)$ can be evaluated using the same numerical machinery used for the $J_{\alpha\beta l}(k)$ integrals.
The final result for $J_{[2,-2,0, \text{reg}]}(k)$ is completely analogous to the method in $\S$ \ref{method}, with the only exception that $g_n$ is replaced by $g_n^\text{reg}$ and the factor $2^{Q_h}$ is replaced by $2^{Q_{-2, n}}$ in Eq.~(\ref{J_k_rk}). {\sc FAST-PT} allows the user to specify which case is desired.

\section{Performance}

We now discuss the results from the {\sc FAST-PT} algorithm. Unless otherwise noted, results are based on the input linear power spectrum generated by the Boltzman solver CAMB \citep{2000ApJ...538..473L}, assuming a flat $\Lambda$CDM cosmology corresponding to the recent Planck results \cite{2015arXiv150201589P}. Timing results were obtained on a MacBook Pro Retina laptop computer, with a 2.5 GHz Intel Core i5 processor and running OS X version 10.10.3. We used Python version 2.7.10, {\sc numpy} 1.8.2, and {\sc scipy} 0.15.1.

\subsection{1-loop Results}

\begin{figure}[tbp]
\centering 
\includegraphics[width=.9\textwidth]{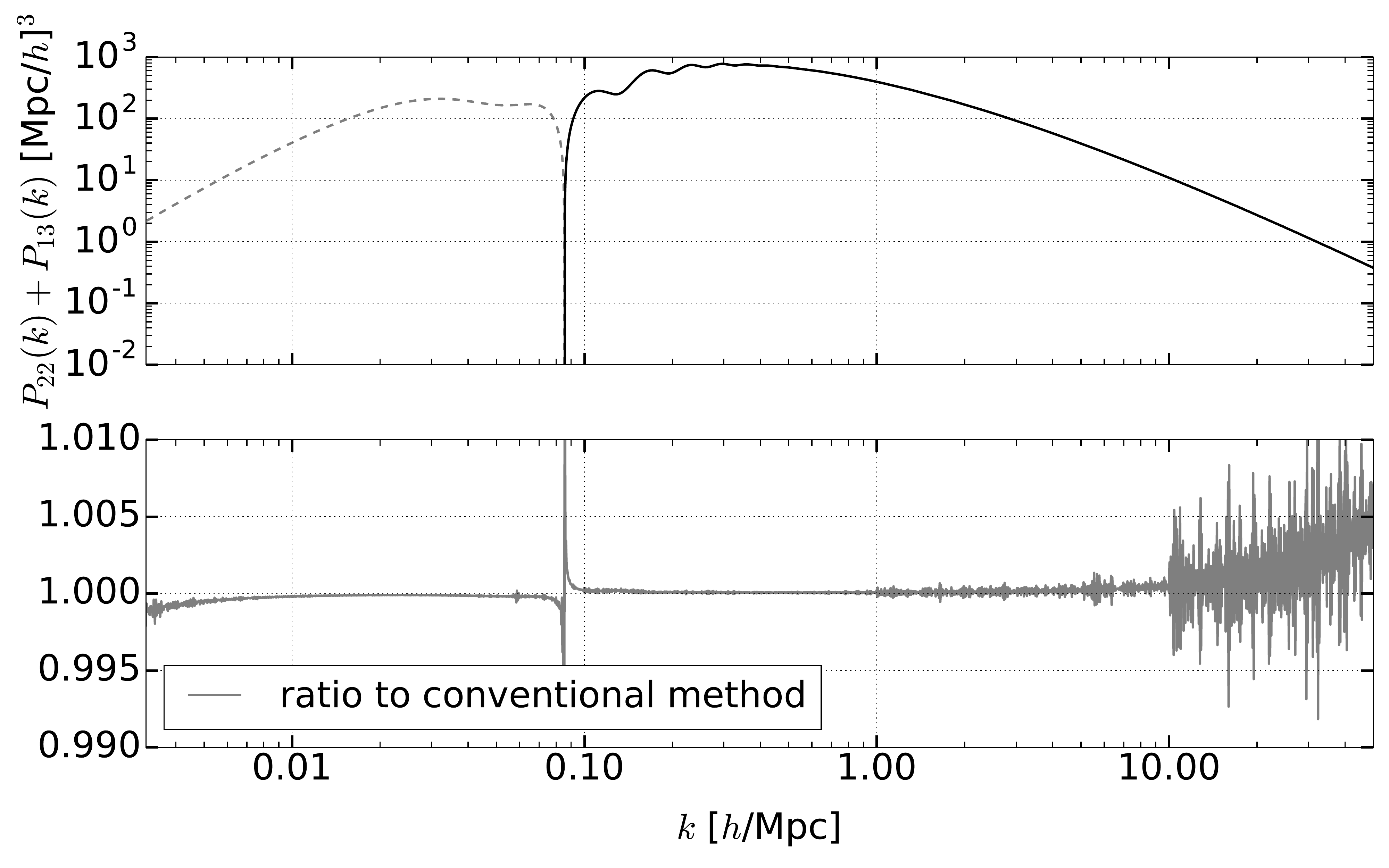}
\caption{\label{fig:result} {\sc FAST-PT} 1-loop power spectrum results versus those computed using a conventional fixed-grid method. The top panel shows {\sc FAST-PT} results for $P_{22}(k) + P_{13}(k)$ (the dashed line is for negative values). The bottom panel plots the ratio between {\sc FAST-PT} and the conventional method.}
\end{figure}

To test our method we evaluated the 1-loop SPT correction to the power spectrum, $P_{22}(k) + P_{13}(k)$. We sample the power spectrum for 3000 $k$-points from $\log_{10} k_\text{min}=-4$ to $\log_{10} k_\text{max}=2$ and we additionally pad our input signal with 500 zeros at both ends of the array. A typical run for a sample of this size takes {\sc FAST-PT} a total time $\sim 0.02 $ seconds on a laptop.
We recommend that {\sc FAST-PT} users sample the input power spectrum on a grid larger than desired and then trim the output to the desired range to avoid wrapping effects. We take this approach and present our results on a grid from $  k_\text{min}=0.003$ to $ k_\text{max}=50$. The top panel Fig.~\ref{fig:result} plots our {\sc FAST-PT} results, while the bottom panel plots the ratio of our {\sc FAST-PT} calculations to a conventional method.\footnote{The ``conventional'' method is a fixed-grid 2D integration code. Here $P_{22}$ was computed by putting ${\mathbf k}$ on the $z$-axis and writing ${\mathbf q}$ in cylindrical coordinates. The azimuthal integral is trivial. We sample the integrand logarithmically in the radial direction $q_\perp$, and stretch the vertical direction according to $q_z/k = 1+\sinh(20\upsilon)/[2\sinh(20)]$, with $\upsilon>-1$. This samples half of space (so the result must be doubled) and by uniformly sampling in $\upsilon$, it places higher resolution near ${\mathbf q}\approx{\mathbf k}$, which is important to correctly sample the contribution to $P_{22}$ from advection by very long-wavelength modes. The $P_{13}$ integral was log-sampled in $r$.}
We observe that the 1-loop power spectrum{\sc FAST-PT} agrees with the conventional method to high precision. The noise observed in the bottom panel of Fig.~\ref{fig:result} is due to noise in the input power spectrum from CAMB; any integration method must interpolate this noise, and this results in noise in the output spectrum $P_{22}+P_{13}$ which differs depending on the method. At high $k$, the noise in $P_{22}+P_{13}$ is larger than (and of opposite sign to) the noise in $P_{\rm lin}(k)$, which is a phenomenon common to diffusion problems and is the correct mathematical solution to SPT, where re-normalization or re-summation techniques are not used (see \S\ref{sec:RG}). The sharp spike around $k=0.1 h$/Mpc is due to the zero crossing of the 1-loop power spectrum, where ratios of corrections suffer from a ``0/0'' ambiguity. We conclude that differences between {\sc FAST-PT} results and those from our conventional method are negligible on the scales of interest.

\begin{figure}[tbp]
\centering 
\includegraphics[width=.9\textwidth]{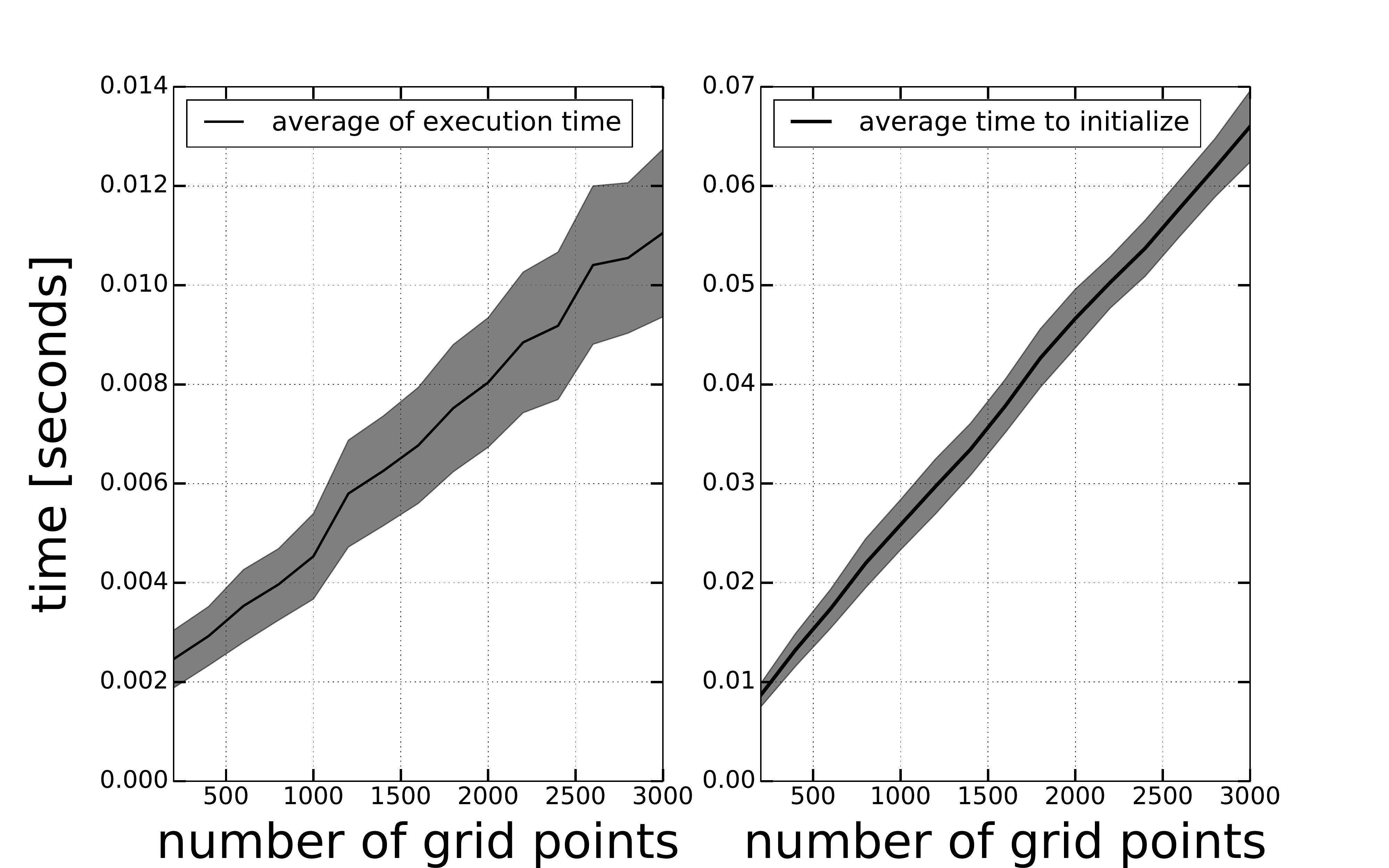}
\caption{\label{fig:time_plot} Estimate of {\sc FAST-PT} execution time to number of grid points scaling. The left panel plots the average one-loop evaluation time, after initialization of the {\sc FAST-PT} class. The right panel plots the average time required for initialization of {\sc FAST-PT} class for 1500 runs. For a sample of grid points, the error is computed by taking the standard deviation of 1500 runs.}
\end{figure}

Fig.~\ref{fig:time_plot} plots estimated run time versus grid size. A solid black line in the left panel plots the average recurring time (i.e.\ the time of execution after initialization of the {\sc FAST-PT} class) for 1500 runs. The grey band covers the area enclosed by $\pm$ one standard deviation. The right panel plots the average initialization time for 1500 runs, i.e.\ the time to initialize the {\sc FAST-PT} python-class and evaluate all functions that only depend on grid size (for example $g_{\alpha n}$). The total time for one one-loop evaluation is the addition of the black line in the right and left panels. Run time can vary across machines, so Fig.~\ref{fig:time_plot} serves only as an estimate. 

\subsection{Renormalization Group Flow}
\label{sec:RG}

The renormalization group (RGPT) method of  \cite{2007PhRvD..75d3514M,2014arXiv1403.7235M} provides a more accurate model for the power spectrum than SPT \citep{2009PhRvD..80d3531C,2009MNRAS.397.1275W,2011PhRvD..84f3501O}, providing significant improvement to both the structure of the BAO feature and the broadband power at smaller scales (higher $k$). The RG evolution equation is 
\begin{align}
\label{RG_flow}
\frac{ d P(k,\lambda)}{d \lambda} = G_2[ P(k,\lambda), P(k,\lambda)]~, 
\end{align} 
where $G_2[P,P]$ is the standard 1-loop correction to the power spectrum, i.e.\ $P_{22}(k) + P_{13}(k)$ with the caveat that the input power spectrum need not be the linear power spectrum. The parameter $\lambda$ is a ``coupling'' strength parameter proportional to the growth factor squared. One can imagine that Eq.~(\ref{RG_flow}) represents a time-evolution of the power spectrum (in an Einstein-deSitter universe) starting at $P(k, \lambda=0)= \Pl(k)$, moving forward in time by a small step, using perturbation theory to update the power spectrum, and then using the updated power spectrum as the initial condition for the next step, iterating until one reaches $\lambda =1$.

However, despite the potential advantages of the RG approach, it can be quite numerically intensive. Eq.~(\ref{RG_flow}) is a stiff equation and becomes unstable when the integration step size is too large, and it requires an evaluation of the 1-loop SPT kernel at every step. Conventional computational methods are thus extremely time consuming. The speed of {\sc FAST-PT} makes this calculation significantly more feasible. We have compared our RG flow results with those obtained from the {\sc Copter} code \cite{2009PhRvD..80d3531C,2013ascl.soft04022C}, a publicly available code written in C++. We have found that for RG flow our code can obtain results in substantially less time. For instance, on a 200 point grid, from $k_\text{min}=0.01$ to $k_\text{max}=10$, our {\sc FAST-PT} RG flow results take $\sim 5$ seconds, while {\sc Copter} RG flow results take over 5 minutes. In Appendix \ref{rg_int} we explain our integration routine, as well as document RG-flow run times for various grid sizes. A {\sc FAST-PT} user must consider the stiff nature of Eq.~(\ref{RG_flow}) when choosing a step size for the integration; we recommend that they consult Appendix \ref{rg_int}.

\begin{figure}[tbp]
\centering 
\includegraphics[width=.9\textwidth]{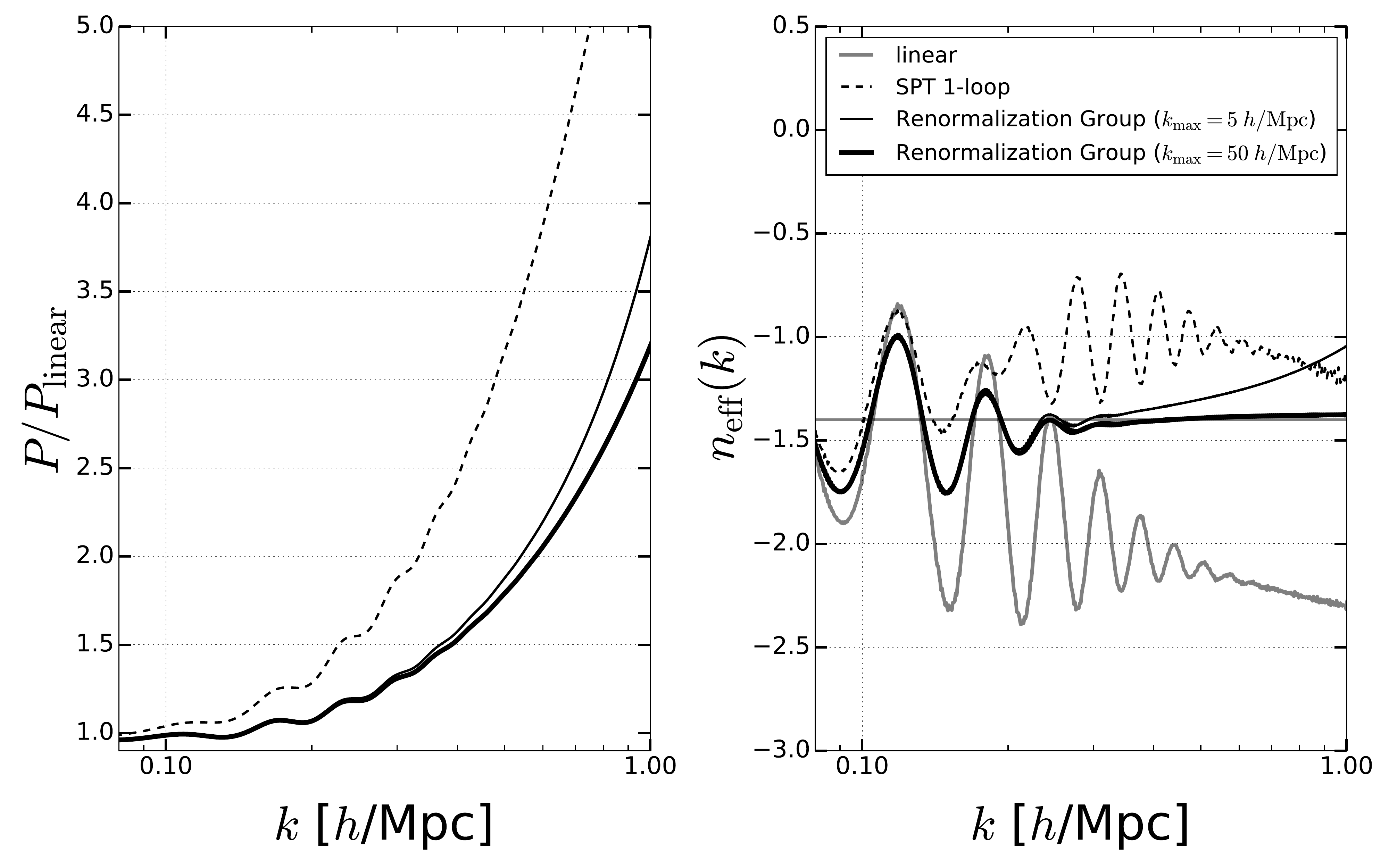}
\caption{\label{fig:ii} {\sc FAST-PT} Renormalization group results for $k_\text{max}=\{5, 50 \} h \Mpc^{-1}$. Left panel shows Renormalization group results and SPT results compared to the linear power spectrum (see legend in right panel). Right panel shows $n_\text{eff}= d \log P/d\log k$ for Renormalization group, SPT, and linear theory.}
\end{figure}

The left hand panel of Fig.~\ref{fig:ii} shows our renormalization group and SPT results compared to linear theory. In our analysis we performed two renormalization group runs: one to $k_\text{max}=5\; h \Mpc^{-1}$ and another to $k_\text{max}=50\; h \Mpc^{-1}$. Our results are consistent with the plots found in \cite{2007PhRvD..75d3514M} (note that in our runs we include the BAO feature). The right hand panel of Fig.~\ref{fig:ii} plots the effective power law index as a function of $k$,  $n_\text{eff}= d \log P/d\log k$. Here we see two characteristic features of RG evolution, the damping of the BAO and  $n_\text{eff}$ approaching a fixed point value of $\sim -1.4$. 

\begin{figure}[tbp]
\centering 
\includegraphics[width=.9\textwidth]{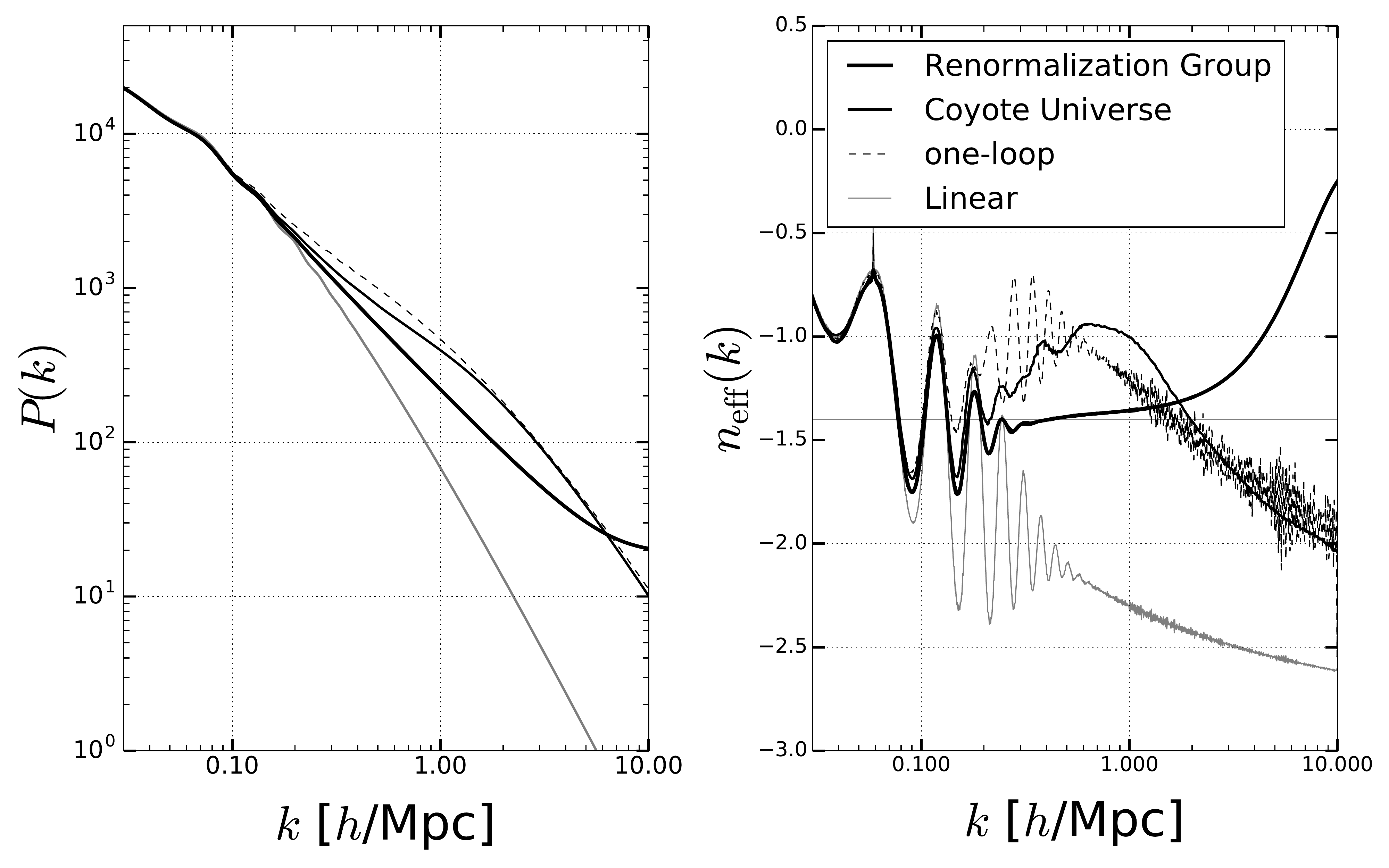}
\caption{\label{fig:rg_plateau} Renormalization group results compared to standard 1-loop calculations and those taken from the {\sc Coyote Universe}.  Left panel plots power spectra. A plateau at high-$k$ develops due to boundary conditions. Right panel shows $n_\text{eff}(k)= d \log P/d\log k$.}
\end{figure}

Figure~\ref{fig:rg_plateau} shows the effect of a boundary condition within our numerical algorithm. We integrate Eq.~(\ref{RG_flow}) for some $k_\text{min}$ and $k_\text{max}$. The $k_\text{max}$ boundary does not allow for power to continuously flow from larger to smaller scales, as would occur for infinite boundary conditions. As a result power builds up at high-$k$ causing the plateau observed in left panel of Fig.~\ref{fig:rg_plateau}. The right panel of Fig.~\ref{fig:rg_plateau} shows $n_\text{neff}$. We do see that before the onset of the plateau $n_\text{eff}$ does approach $\sim -1.4$ and this designates a region where RG results at finite $k_{\rm max}$ reproduce the asymptotic behavior as $k_{\rm max}\rightarrow\infty$. 

To qualify the accuracy of the RG method in the weakly non-linear regime, we also plot results from the {\sc FrakenEmu} emulator (in Fig~\ref{fig:rg_plateau}), which is based of the {\sc Coyote Universe} simulations \cite{2014ApJ...780..111H}. In the vicinity of $k\sim 0.1$, it is observed that RG methods better follow the fully non-linear results of the {\sc Coyote Universe}. 

Figure \ref{fig:rg_plateau} also shows another interesting feature, the removal of noise in the RG-framework. As mentioned earlier, linear power spectrum generated by CAMB contains low-level noise. This noise is most easily visualized through a derivative, for instance $n_\text{eff}$. One can see that $n_\text{eff}$ for the linear power spectrum in Fig.~\ref{fig:rg_plateau} is noisy, particularly at large $k$. Under the RG evolution this noise is washed away, as seen in the RG $n_\text{eff}$ results. This is a result of the fact that noise in $\Pl(k)$ results in ``negative'' noise features in $P_{13}(k)$. Under the RG flow, this feature causes noise initially present to be smeared away in the nonlinear regime. This is also what happens in the real universe, since features in the power spectrum at small $\Delta k$ correspond to correlations at large real-space scales $\sim 2\pi/\Delta k$, which are smeared out by advection; this effect is responsible for the familiar BAO peak smearing \cite{2007ApJ...665...14S}.

\section{Summary}

In this paper we have introduced {\sc FAST-PT}, an algorithm (and associated public code) that quickly evaluates convolution integrals in cosmological perturbation theory. The code is modular and written in a high-level language (Python), and it is extremely fast due to algorithmic improvements. The keys to the method are locality (expressing the Fourier-space mode coupling integrals $J_{\alpha\beta l}$ as a product of correlation functions in configuration space); scale independence of the physics of gravity ("hence the utility of a power-law decomposition for the power spectrum); and the FFT (which enables log-spaced data to be converted into a superposition of power laws and vice versa). The recurring cost of the 1-loop SPT calculations is presented in Fig.~\ref{fig:time_plot}; for a linear power spectrum sampled on a 3000-point grid, one can expect to obtain results in $\sim 0.01 $ seconds. The time for RG results in tabulated in Tables \ref{tab:RG_stab_RK4} and \ref{tab:RG_stab_STS}. For a linear power spectrum sampled on 500-point grid from $k_\text{min}=0.001$ to $k_\text{max}=10$, RG results are obtained in a few seconds. 

We have demonstrated {\sc FAST-PT} in the context of 1-loop SPT and the RG flow. However, similar convolution integrals appear in numerous contexts, making both the conceptual improvements behind {\sc FAST-PT} and the code itself an efficient and flexible tool for the community. For instance, the structure of the 1-loop SPT calculation contains all elements necessary for nonlinear biasing in the galaxy-galaxy and galaxy-matter power spectra. Furthermore, in follow-up work we are extending the technique to other problems in cosmological perturbation theory. In \cite{Fang2016}, we generalize {\sc FAST-PT} to ``tensor'' quantities (broadly defined as those that with explicit dependence on the line-of-sight), including those relevant to the intrinsic alignments of galaxies (e.g.\ \cite{2004PhRvD..70f3526H,2015JCAP...08..015B}), redshift space distortions, and CMB anisotropies. We are additionally exploring further applications of the {\sc FAST-PT} framework. For instance, when the evolution of fluctuation modes is given by a scale-dependent propagator, the time- and scale-dependence of each mode can no longer be separated \cite{2011JCAP...10..037A}. Such a scenario arises in the presence of massive neutrinos, where growth of structure is suppressed on small scales due to free-streaming \cite{2008PhRvL.100s1301S,2009PhRvD..80h3528S,2014JCAP...11..039B}. Solving for nonlinear evolution in such a scenario can be done using a time-flow approach \cite{2008JCAP...10..036P}, requiring many evaluations of mode-coupling integrals. It is similar to the RG flow described above, but with additional complications, particularly due to the scale dependence of the propagators (note that our $J_{\alpha\beta l}$ integrals include only power-law dependences on the magnitudes of $q_1$ and $q_2$), and the fact that the Green's function solution for the bispectrum (needed to reduce the power spectrum solution to a mode-coupling integral) involves products of power spectra at unequal times. We are investigating the extent to which these issues can be treated in {\sc FAST-PT}. Additionally we are exploring the applicability of the {\sc FAST-PT} method to 2-loop calculations. Fast methods to compute the power spectrum past 1-loop order already exist \cite{2012MNRAS.427.2537C, 2012PhRvD..86j3528T}. These methods rely on multi-point propagator techniques. We are working to determine whether FAST-PT-like algorithms can be extended to the 2-loop convolution integrals with computation time comparable to that obtained here for the 1-loop case.

The value of {\sc FAST-PT} lies in its short execution time and the general applicability of these mode coupling integrals to cosmological observables.
Additionally the modular structure of {\sc FAST-PT} makes it easily integrable into cosmological analysis projects, for example those found in \cite{2014MNRAS.440.1379E,2015A&C....12...45Z,2016arXiv160105779K}. Our Python code is publicly available at {\tt https://github.com/JoeMcEwen/FAST-PT} and includes a user manual. We also provide Python scripts to reproduce 1-loop power spectrum, galaxy bias power spectrum, renormalization group results, and animations for renormalization group results.

\acknowledgments
JM is supported by NSF grant AST1516997. XF is supported by the Simons Foundation, and CH by the Simons Foundation, the US Department of Energy, the Packard Foundation, and NASA. JB is supported by a CCAPP Fellowship. JM thanks Ben Wibking for useful discussions concerning the numerical integration of stiff equations and Chris Orban for help setting up {\sc Copter}. We thank David Weinberg for general feedback on the {\sc FAST-PT} project.

\bibliographystyle{JHEP.bst}
\bibliography{refs}

\appendix
\section{Mathematical Identities}
In this work we have used a number of common mathematical identities. These identities are easily found in any standard mathematical physics text, (e.g.\ \cite{abramowitz1964handbook}). However, to make our paper self-contained we list those relevant to our paper. In $\S$ \ref{method} we used the following special function identities:
the addition theorem
\begin{align}
\label{add_form}
P_l( \hat{\mathbf q}_1 \cdot \hat{\mathbf q}_2) =\frac{4 \pi}{2l + 1} \displaystyle \sum_{m=-l}^l Y_{lm}(\hat{\mathbf q}_1) Y^\ast_{lm}(\hat{\mathbf q}_2);
\end{align}
the special case thereof,
\begin{align}
\label{sph_ortho}
\displaystyle \sum_{m=-l}^l Y_{lm}(\theta, \phi)Y^\ast_{lm}(\theta, \phi) = \frac{2l + 1}{4 \pi} ~;
\end{align}
the orthonormality relation
\begin{align} 
\int_{S^2} d^2 \hat{\mathbf k}\, Y_{lm}(\hat{\mathbf k}) Y_{l'm'}^\ast(\hat{\mathbf k}) = \delta_{l l'}\delta_{mm'} ~;
\end{align} 
and the expansion/decomposition of a plane wave:
\begin{align}
\label{angl_1}
\int_{S^2} d^2 \hat{\mathbf q}\, Y^\ast_{lm}(\hat{\mathbf q}) e^{i \mathbf{q} \cdot \mathbf{r}} = 4 \pi  i^l j_l(qr) Y^\ast_{lm}(\hat{\mathbf r})
~~~\leftrightarrow~~~
 e^{i \mathbf{q} \cdot \mathbf{r}} = 4 \pi \displaystyle \sum_{l} \displaystyle i^l j_l(qr)  \sum_{m=-l}^l Y^\ast_{lm}(\hat{\mathbf q})Y_{lm}(\hat{\mathbf r})~.
 \end{align}

\section{$\Gamma$-function identities and evaluations}

We make extensive use of the following integral (see pg. 486 of Ref.~\cite{abramowitz1964handbook}):
\begin{align}
\int_0^\infty dt \; t^\kappa J_\mu(t) = 2^\kappa \frac{ \Gamma \left[ (\mu + \kappa + 1)/2 \right]}{ \Gamma \left[ (\mu - \kappa + 1)/2 \right]}= 2^\kappa g(\mu,\kappa)~, \;\;\; \Re \kappa < 1/2~, \;\;\; \Re(\kappa + \mu) > -1~,
\label{TSJ}
\end{align}
where we define the $\Gamma$-function ratio:
\begin{align}
g(\mu, \kappa)= \frac{ \Gamma \left[ (\mu + \kappa + 1)/2 \right]}{ \Gamma \left[ (\mu - \kappa + 1)/2 \right]} ~.
\label{gmdef}
\end{align}

A second useful integral is
\begin{align} 
\label{gamma_sin_eqn}
\begin{split}
f(\rho) \equiv \int_0^\infty dt \,t^{\rho-1} \sin t  = \Gamma(\rho) \sin\frac{\pi \rho}{2} = \frac{\sqrt\pi}2\,2^\rho g\left(\frac12,\rho-\frac12\right) 
\end{split}
\end{align} 
for $-1<\Re \rho<1$. The second equality is Eq.~(3.761.4) of Ref.~\cite{1994tisp.book.....G}. The last expression is an evaluation of the integral Eq.~(\ref{TSJ}) and the relation
\begin{equation}
\sin t = \sqrt{\frac{\pi t} 2}\,J_{1/2}(t).
\end{equation}
We use the second or third expressions to define $f(\rho)$ via analytic continuation to values of $\rho$ outside the domain of convergence of the integral.

The numerical evaluation of $g(\mu,\sigma)$ in {\sc FAST-PT} uses the {\sc scipy} {\tt gamma} function for most values. However, when the argument to the $\Gamma$-function has a large complex value numerical overflows may occur. Therefore when $|\Im\sigma|>200$ we use an asymptotic form for our evaluations [Eq.~(6.1.40) of Ref.~\cite{abramowitz1964handbook}]:
\begin{align}
\log \Gamma(z) \approx (z-1/2) \log z - z + \frac{1}{2} \log 2\pi + \displaystyle \sum_{m=1}^\infty \frac{ B_{2m} }{2m(2m-1)z^{2m-1}  } ~, 
\end{align}
for $ z \to \infty $ in $ | \arg z | < \pi$. Here $B_{2m}$ are the Bernoulli numbers; we find that only the first two terms $B_2=\frac16$ and $B_4=-\frac1{30}$ are necessary at $|\Im\sigma|>200$ to achieve an error of $<10^{-13}$ in $\log\Gamma(z)$. To avoid overflows, the logarithms are differenced to give $\log g(\mu,\kappa)$ and this result is exponentiated.

The function $f$ in Eq.~(\ref{gamma_sin_eqn}) may be numerically evaluated using a $\Gamma$-function routine, but we prefer to use our routines for $g$. This is because $f$ itself is well-behaved near $\rho=0$ (the $\Gamma$ function has a simple pole and the sine function has a single zero), but the $\Gamma$-function expression in Eq.~(\ref{gamma_sin_eqn}) is ill-behaved. In contrast, our implementation of the function $g$ is well-behaved at $\rho=0$.

\label{sec:math_details}

\section{Mitigation of Edge Effects}
\label{sec:edge}

Fourier methods are susceptible to ringing effects due to discontinuities in the input signal.  To mitigate ringing we smoothly tapper the array edges of the Fourier coefficient array $c_m$ with a window function defined to have continuous first and second derivatives: 
\begin{align} 
W(x)=
\begin{cases}
  \frac{x-x_\text{min}}{x_\text{left}-x_\text{min} }  - \frac{1}{2 \pi} \sin \Big( 2 \pi \frac{x-x_\text{min}}{x_\text{left}-x_\text{min} }   \Big) ~ &  x < x_\text{left} \\
 1 &  x_\text{left} < x < x_\text{right} \\
 \frac{x_\text{max}-x}{x_\text{max} - x_\text{right} }  - \frac{1}{2 \pi} \sin \Big( 2 \pi \frac{x_\text{max}-x}{x_\text{max} - x_\text{right} }   \Big)  &  x > x_\text{right}
\end{cases}
~,
\end{align} 
where $x_\text{left}$ and $x_\text{right}$ are input parameters that determine the position of the tapering. For a typical run we dampen the high frequency Fourier modes by applying the window function to $c_m$. In this case the position where the tapering begins is at an $m =\pm 0.75 \times N/2$, where $N$ is the size of the input array. 

Fourier analysis assumes the input signal to be periodic. This often leads a a wrap-around effect in our results, i.e.\ the leakage between low-$k$ to high-$k$. To alleviate this effect we allow for zero-padding of the input power spectrum. For an input power spectrum sampled on a $k$-grid of a few thousand points, we add $\sim 500$ zeros to both ends. Wrap-around effects can also be mitigated by using an input power spectrum sampled over a larger $k$-range than desired and then trimming on output, we recommend one take this approach in combination with filtering of the Fourier coefficients and zero-padding.

\section{RG-flow Integration}
\label{rg_int}
Numerical integration of Eq.~(\ref{RG_flow}) will quickly develop instabilities when using a simple integration routine (e.g.\ Euler integration). These instabilities are highly sensitive to $k_\text{max}$ and the linear grid spacing $\Delta$. 
For $k_\text{max} \leq 1$ we have found that a fourth order Runge-Kutta (RK4) method will produce stable results. However,  for $k_\text{max} >1$, stable RK4 results require an integration step $\Delta \lambda$ greater than $10^{-3}$, decreasing rapidly with increasing $k_\text{max}$. An integration step this small will increase computation time substantially. To decrease computation time we have implemented the super time step (STS) method of \cite{alexiades1996super}. Super time step methods are a class of integrators developed to solve parabolic equations, often for diffusion problems. They belong to the family of Runge-Kutta-Chebyshev methods and have the advantage that they can decrease the computation time by increasing the stability region. For each integration step  $\Delta \lambda$, the STS method takes $N_s$ inner Euler steps $\delta \lambda_j$, where $j=1,2,...,N_s$, such that $\Delta \lambda = \sum_j \delta \lambda_j$.  The $\delta \lambda_j$ are chosen by

\begin{align}
\delta \lambda_j = \Delta \lambda_\text{CFL} \left[ (\mu -1)\cos \frac{\pi(2 j-1)}{2N_s}  + (1 + \mu) \right]^{-1} ~, 
\end{align} 
where $ \Delta \lambda_\text{CFL}$ is the usual Courant-Friedrichs-Lewy stability step, $\mu$ is a damping factor (related to a ratio of eigenvalues) and is between 0 and 1. Equation~(\ref{RG_flow}) is not a parabolic partial differential equation; it is an integro-differential equation, which behaves as a diffusion equation under certain limiting circumstances. As such we do not have a rigorous theory for selecting $\Delta \lambda_\text{CFL}$ and $\mu$, and we chose their values by numerical experiment. For $k_\text{max}=10$ and 2000 grid points we have chosen $ \Delta \lambda_\text{CFL}=0.001$, $\mu=0.1$, and $N_s=10$. A {\sc FAST-PT} user has the option to specify  $ \Delta \lambda_\text{CLF}$, $\mu$, and $N_s$. In tables \ref{tab:RG_stab_RK4} and \ref{tab:RG_stab_STS} we document RG-flow run times for various grid sizes. These tables should serve as guidance when choosing the integration routine and/or routine parameters. 

We also control stability by filtering the right hand side of Eq.~\ref{RG_flow} at each integration step with the window function presented in Appendix~\ref{sec:edge}. The tapering of the window function begins at $\log k_\text{min} +0.2$ and $\log k_\text{max} -0.2$. Applying this window function smooths any sharp features introduced by the edge effects, slowing the development of instabilities due to the stiff nature of the differential equation.

\begin{table}[tbp]
\centering
\begin{tabular}{| l | c | c| r |}
\hline
$\Delta  \log k$ &  grid points & run time [seconds] \\
\hline
0.14 & 50 & 0.24\\
0.069 & 100 & 0.27 \\
0.013  & 500 & 0.30 \\
0.0045  & 1500 & 0.45 \\
\hline
\end{tabular}
\caption{\label{tab:RG_stab_RK4} Stable  RK4 runs for $k_\text{min}=10^{-3}$ and $k_\text{max}=1$ and $\Delta \lambda=0.1$.}
\end{table}

\begin{table}[tbp]
\centering
\begin{tabular}{| l | c | c| r |}
\hline
$\Delta  \log k$ & grid points & run time [seconds] \\
\hline
0.018& 500 & 2.77\\
0.0092 & 1000 & 3.55\\
0.0046 &  2000 & 5.10 \\
\hline
\end{tabular}
\caption{\label{tab:RG_stab_STS} Stable STS runs for $k_\text{min}=10^{-3}$ and $k_\text{max}=10$. Results were obtained using STS parameters: $\mu=0.1,~ \Delta \lambda_\text{CFL}=0.001,~N_s=10$.}
\end{table}

\end{document}